# Identification and Removal of System-Induced Autofluorescence in Miniaturized Fiber-optic Fluorescence Endoscopes

Lei Xiang[1, 2], Rouyan Chen[1, 2, 4], Joanne Tan[3, 4], Victoria Nankivell[3, 4], Christina A. Bursill[3, 4], Robert A. McLaughlin[2, 3], Jiawen Li[1, 2, *]

[1] School of Electrical and Mechanical Engineering, University of Adelaide, Adelaide, SA 5005, Australia

[2] Institute for Photonics and Advanced Sensing, University of Adelaide, Adelaide, SA 5005, Australia

[3] Faculty of Health and Medical Sciences, University of Adelaide, Adelaide, SA 5005, Australia

[4] Lifelong Health Theme, South Australian Health and Medical Research Institute (SAHMRI), Adelaide, SA 5000, Australia

* To whom correspondence should be addressed:
**Email:** jiawen.li01@adelaide.edu.au

**Author Contributions:**

Conceptualization: LX, JL
Formal Analysis: LX, RC
Funding Acquisition: JL, CAB, RAM
Investigation: LX, RC, JT, VN, CAB, JL
Methodology: LX, CAB, RAM, JL
Resources: CAB, RAM, JL
Software: LX
Supervision: RAM, JL
Validation: LX, JT, JL
Visualization: LX, JT, JL
Writing – Original Draft Preparation: LX
Writing – Review & Editing: LX, RC, JT, VN, CAB, RAM, JL

**Competing Interest:**

**Classification:** Physical Sciences: Engineering

**Abstract**

Miniaturized fiber-optic fluorescence endoscopes play a crucial role in medical diagnostics and research, but system-induced autofluorescence remains a significant challenge, particularly in single-fiber setups. While recent advances, such as double-clad fiber (DCF) and DCF couplers, have reduced background noise, complete elimination remains challenging. Research on the various sources of system-induced autofluorescence and the methods to remove them are scarce.

This study seeks to fulfill this need by proposing practical approaches to removal of system-induced autofluorescence. This study presents the methods to suppress static background noise and proposes an algorithm based on least-squares linear spectral unmixing to remove variable system-induced autofluorescence artifacts. The algorithm was evaluated on a single-fiber DCF intravascular imaging system, with phantom and rodent *in vivo* experiments confirming its effectiveness. Results showed accurate differentiation between true sample fluorescence and system-induced autofluorescence artifacts through the validation with optical coherence tomography images and histology results, further verified by statistical analysis. Unlike simple background subtraction, the method addresses dynamic variations in background noise and incidental artifacts, providing robust performance under varying conditions. Our method may be adapted to various fiber-based endoscopy setups and be compatible with different fluorescent agents and autofluorescence imaging, broadening its applicability in biomedical imaging.

**Significance Statement**

System-induced autofluorescence presents a persistent challenge in miniaturized fiber-optic fluorescence endoscopes, often confounding the accurate detection of sample fluorescence signals. This issue is particularly pronounced in single-fiber setups where space constraints limit traditional methods of noise suppression. By investigating the various sources of system-induced autofluorescence and introducing a novel least-squares linear spectral unmixing algorithm, this study offers a robust solution to differentiate and eliminate system-induced autofluorescence. The proposed algorithm is adaptable across various system setup and fluorophores and holds potential for broader applications, advancing the precision of fiber-optic fluorescence endoscopy in clinical and research settings.

**Introduction**

Miniaturized fiber-optic fluorescence endoscopy has become an essential tool in medical diagnostics and research, offering high-resolution visualization of molecular information in biological tissue. This technology is extensively used in cardiology [1, 2], gastroenterology [3, 4], oncology [5, 6], and pulmonology [7, 8], where it plays an important role in the precise detection and diagnosis of diseases. Fiber-optic endoscopic probes can be classified into two broad configurations: multiple-fiber setups, which separate excitation and collection channels to reduce background autofluorescence signals; and single-fiber setups, which combine illumination and collection in a more compact and simplified design. The latter is particularly advantageous for imaging in small blood vessels and airways due to its minimal size and straightforward configuration [9, 10].

In the single-fiber setup, system-induced autofluorescence is a significant confounding issue that limits the sensitivity and measurement accuracy of these systems. Noise in the measured fluorescence signal originates from two main sources: a static background autofluorescence from excitation light traveling through the system, and variable autofluorescence artifacts caused by reflected excitation light from external hyperreflective objects [11, 12]. The first source (background noise) usually remains constant during scanning of a sample. This arises from autofluorescence generated by the excitation light as it passes through the system components and out toward the sample. This commonly arises from the photoactive compounds in the fiber, particularly the doping material in the fiber core [12]. Side-band components of the excitation light, where they overlap with the measured fluorescence wavelengths, can also contribute to this noise if an appropriate bandpass filter is not present to clean-up the excitation spectrum [13].

To reduce the measured autofluorescence generated as the excitation light passes through the fiber, one recent implementation of a single-fiber system used double-clad fibers (DCF) and DCF couplers [7, 12, 14]. In this setup, the excitation light travels through the core of the DCF, while sample fluorescence signals are collected through the larger-diameter inner cladding and guided to a multimode fiber via a specially designed DCF coupler. Separation of excitation and collection paths helps reduce the autofluorescence generated from fiber doping material by the excitation light. However, despite these improvements, eliminating background autofluorescence remains a challenge due to mis-coupling of the excitation laser from the fiber core into the inner cladding [12].

A second source of noise in the measured fluorescence arises from incidental artifacts that change with the sample being scanned. This can originate from excitation light that is reflected back into the fiber-probe from external hyperreflective surfaces [15] such as a catheter, or compositions in the biological tissue (e.g., crystals, calcifications). This type of noise will vary within a scan and cannot be mitigated by subtracting a constant background signal. It can lead to false positives in detection of tissue fluorescence when there is specular reflection. In multifiber systems, it may be possible to affix a filter to the detection fiber to reduce this component of noise. In practice, it is often challenging to use such a filter because of the highly miniaturized space-constraints in miniaturized endoscopic probes. In a single-fiber system this is not possible as it would block excitation light from illuminating the sample. There is a need to find new approaches to mitigate this class of noise in the measured fluorescence signal.

To the best our knowledge, there is limited study on the various sources of system-induced autofluorescence and the methods to remove them, whether constant during scanning or dynamic with the sample. This study aims to address this need and propose an integrated software and hardware solution that offers a straightforward and practical approach to removal of system-induced autofluorescence.

In this paper, we first summarize the methods to suppress the static system-induced autofluorescence noise, which is critical to improve system sensitivity and signal-to-noise ratio. We then propose an algorithm utilizing least-squares linear spectral unmixing, which can automatically estimate system-induced autofluorescence noise and sample fluorescence signal. This approach is implemented using a single-fiber DCF intravascular imaging system. Our method is then applied to phantom data, and *in vivo* data of mouse arteries. The results showed that our method effectively mitigates both static background noise and dynamic sample-based autofluorescence artifacts.

**Theory**

**1. Spectral Unmixing**

Typically, the static background noise in fluorescence endoscopy can be addressed by subtracting a pre-acquired background signal. However, simple background subtraction proves ineffective in addressing dynamic fluorescence artifacts caused by incidental specular reflection and elastically scattered light that is coupled back into the fiber probe and induces additional system auto-fluorescence. To overcome these limitations, we have developed a method focused on post-processing the detected fluorescence emission spectrum to eliminate variable system-induced autofluorescence. Our method is based on spectral unmixing, a widely used technique across various fields including remote sensing, environmental monitoring, and material classification [16]. Several unmixing methods have been developed to handle different types of data and noise profiles. Among these methods, least-squares linear unmixing is computationally efficient but struggles when spectral profiles of the components are unknown [16]. As a blind source separation method, independent component analysis operates under the assumption that the source signals are statistically independent and non-Gaussian [17]. However, in real-world scenarios, these assumptions may not always hold true, potentially leading to results that lack physical meaning or interpretability, especially when negative sources are extracted. Non-negative matrix factorization (NMF) is another blind source separation method, which is highly effective for complex, mixed signals without predefined profiles but is computationally demanding and prone to non-unique solutions [18, 19]. Selecting appropriate sparsity and geometric constraints is crucial to ensuring unique and interpretable results. These constraints need to be tailored carefully to the specific characteristics of the tissue and the imaging system to maximize the accuracy and clarity of the unmixing process, while still maintaining the biological relevance of the results.

In the indocyanine green (ICG)-enhanced intravascular near-infrared fluorescence (NIRF) imaging system we use to verify our methods, the autofluorescence of biological tissues in the near-infrared range (700-900 nm) is much lower than fluorescence emitted at visible wavelengths and is also much weaker than the fluorescence emitted by ICG [20, 21]. For our application, this allows us to simplify the spectral analysis by considering only two key spectral components: system-induced autofluorescence and ICG fluorescence. Linear unmixing is a reliable and fast approach for separating these two spectra, provided both spectral signatures, also known as endmembers, are well defined. While the system-induced autofluorescence endmember can be easily measured before *in vivo* imaging, the ICG fluorescence endmember is more challenging to characterize due to *in vivo* spectral shifts [22]. To address this, we employed Pearson correlation to automatically estimate the ICG endmember, enabling the application of linear unmixing for accurate separation of ICG fluorescence from system-induced autofluorescence during imaging.

**2. System Light Path**

Fig. 1 illustrates a representative light path in a single-fiber fluorescence endoscope. Excitation light from the laser source is first filtered using a bandpass excitation filter before being transmitted through the core of the DCF arm of a DCF coupler, and then reaches the stationary side of a fiber optic rotary joint (FORJ). The FORJ then couples the excitation light to the rotating side which is connected to the fiber probe. Upon reaching the target, such as a blood vessel, the probe collects both fluorescence components and reflected incident light from the sample. These components are transmitted through the inner cladding of the DCF and coupled into a multimode fiber. During this process, the reflected excitation light induces autofluorescence within the system, which adds to the fluorescence signals from the sample. A long pass emission filter is applied before the detector to remove the reflected excitation light, though it cannot eliminate system-induced autofluorescence that overlaps with the target fluorescence.

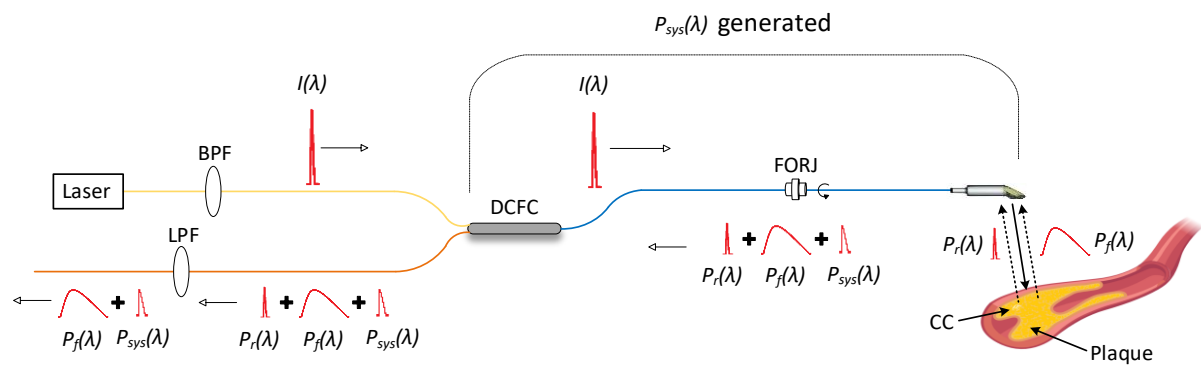

Fig. 1. A representative light path of a single-fiber fluorescence endoscope imaging of atherosclerotic plaques. BPF, bandpass filter; LPF, long pass filter; yellow line, single mode fiber; orange line, multimode fiber; blue line, double-clad fiber; DCFC, double-clad fiber coupler; FORJ, fiber optic rotary joint; CC, cholesterol crystal; $I(\lambda)$, incident light; $P_r(\lambda)$, reflective radiance; $P_f(\lambda)$, sample fluorescent radiance; $P_{sys}(\lambda)$, system-induced autofluorescence.

### 3. Signal models for linear spectral unmixing

In this section, we will demonstrate that the measured spectra can be represented as a linear combination of two constituent spectra: system-induced autofluorescence and sample fluorescence.

For surfaces where the observed reflective radiance $P_r(\lambda)$ of an object only depends on the incident light and its reflectance, it can be described as

$$P_r(\lambda) = I(\lambda)R(\lambda) \qquad (1)$$

where $I(\lambda)$ is the intensity of the incident light and $R(\lambda)$ is the object's reflectance at wavelength $\lambda$, which is a constant for Lambertian surfaces and angle related for specular surfaces.

Assuming narrow-band excitation light, the fluorescent radiance $P_f(\lambda)$ from a pure fluorescent object can be described as the function of the illuminant, the object's excitation, and the emission spectra [23] as shown in Equation (2).

$$P_f(\lambda) = S_{em}(\lambda) \int I(\lambda_i) S_{ex}(\lambda_i) d\lambda_i \qquad (2)$$

where $S_{ex}$ denotes the fluorophore's excitation spectrum, $S_{em}$ is the emission spectrum and $\lambda_i$ is within the range of excitation spectrum.

From Equation (1) and (2), we can express the system-induced autofluorescence generated by the reflected incident light $P_r(\lambda)$ as

$$P_{sys}(\lambda) = S_{em\_sys}(\lambda) \int I(\lambda_i) R(\lambda_i) S_{ex\_sys}(\lambda_i) d\lambda_i \qquad (3)$$

where $S_{ex\_sys}$ and $S_{em\_sys}$ are the system-induced autofluorescence excitation and emission spectra, respectively.

The total spectrum $P(\lambda)$ is the combination of $P_r(\lambda)$, $P_{sys}(\lambda)$ and $P_f(\lambda)$,

$$P(\lambda) = I(\lambda)R(\lambda) + S_{em\_sys}(\lambda) \int I(\lambda_i) R(\lambda_i) S_{ex\_sys}(\lambda_i) d\lambda_i + S_{em}(\lambda) \int I(\lambda_i) S_{ex}(\lambda_i) d\lambda_i \quad (4)$$

After passing the long pass filter, the reflective radiance $P_r(\lambda)$ is removed. For a given narrow band laser excitation, a certain type of fluorophore, and a pre-defined imaging system, $\int I(\lambda_i) R(\lambda_i) S_{ex\_sys}(\lambda_i) d\lambda_i$ and $\int I(\lambda_i) S_{ex}(\lambda_i) d\lambda_i$ are independent of their relative emission spectra, so the Equation 4 can be written as

$$P(\lambda) = \beta_{sys} S_{em\_sys}(\lambda) + \beta_f S_{em}(\lambda) \tag{5}$$

where $\beta_{sys} = \int I(\lambda_i) R(\lambda_i) S_{ex\_sys}(\lambda_i) d\lambda_i$ and $\beta_f = \int I(\lambda_i) S_{ex}(\lambda_i) d\lambda_i$ are scale factors.

From Equation (5) we know that the final observed spectrum is the linear combination of two weighted endmembers: system-induced autofluorescence spectrum and the sample's emission spectrum. This provides us with the possibility to separate them with the least-squares linear unmixing method. We add an error term $\varepsilon$ to Equation 5 to account for any negligible spectral components not from these two endmembers, resulting in Equation 6:

$$P_m(\lambda) = \beta_{sys} S_{em\_sys}(\lambda) + \beta_f S_{em}(\lambda) + \varepsilon \tag{6}$$

The goal is to find the coefficient $\beta_{sys}$ and $\beta_f$ by minimizing $\sum_\lambda \varepsilon^2$, the squared difference between the observed spectra $P(\lambda)$ and modelled spectra $P_m(\lambda)$ over all wavelengths $\lambda$.

**Methods**

**1. Background noise reduction**

Minimizing the background noise before it reaches the detector is important for enhancing the dynamic range and sensitivity of the fiber-optic fluorescence endoscope system. As illustrated in Fig. 1, four main components in a fluorescence endoscope can contribute to system-induced autofluorescence: the DCF coupler, the FORJ, the fiber patch cable between the DCF coupler and the FORJ, and the fiber probe. Antireflection coating is commonly applied to FORJs to reduce reflection and improve return loss [24], which in turn helps lower system autofluorescence. Similarly, new techniques like asymmetric fusion-tapering is employed in

DCF coupler to improve transfer efficiency and reduce excitation light crosstalk [25], thereby minimizing autofluorescence. In this study, we focus on demonstrating the excitation/emission filters to reduce unwanted light and minimizing the fiber length to limit autofluorescence generation along the optical path.

## 2. Unmixing Algorithm

Fig. 2 shows the flowchart of the unmixing algorithm. To be able to use linear unmixing, two endmembers $S_{em\_sys}(\lambda)$ and $S_{em}(\lambda)$ need to be extracted from the combined fluorescence spectrum.

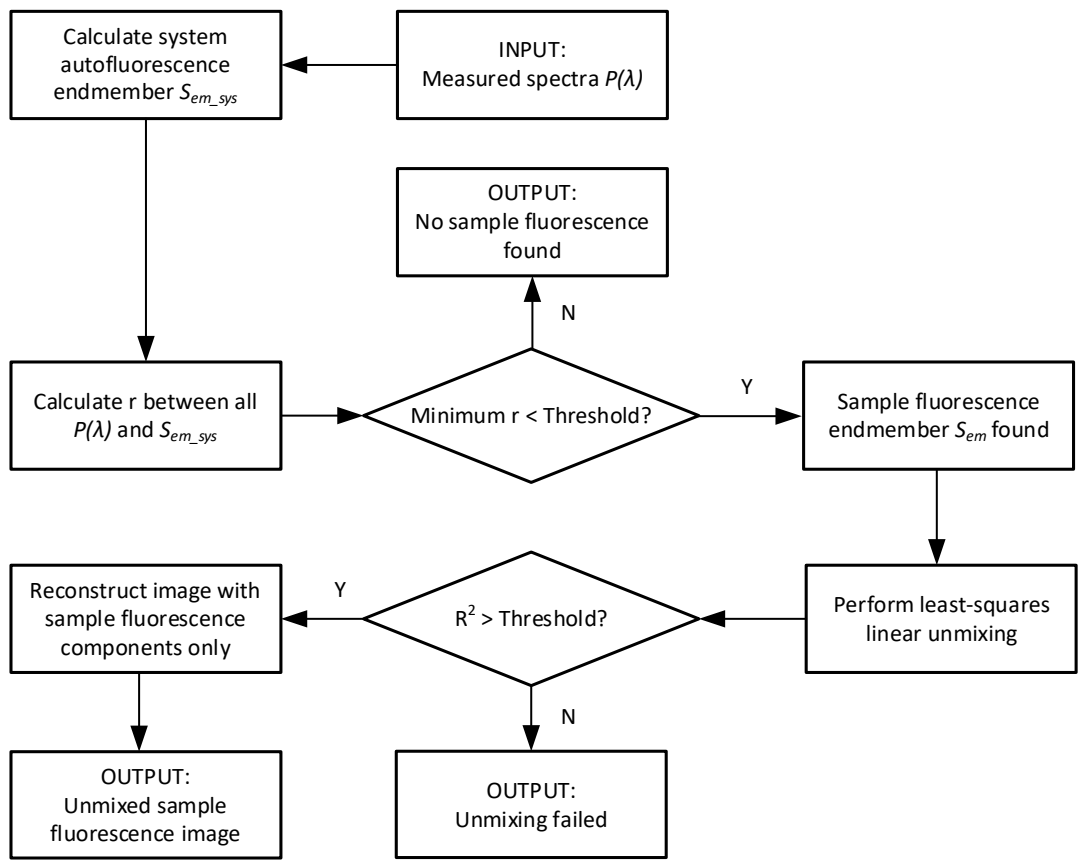

Fig. 2. Diagram of the unmixing algorithm. r, correlation coefficient; $R^2$, coefficient of determination; $P(\lambda)$, measured spectra; $S_{em\_sys}$, system-induced autofluorescence endmember, $S_{em}$, sample fluorescence endmember.

In a system with a single fiber for both excitation and emission, the system-induced autofluorescence endmember spectrum can be simply measured with a specular surface such as a mirror. When working on datasets without a pre-measured autofluorescence spectrum, we might assume that the measured spectrum with the lowest accumulated intensity over the working wavelength consists only the minimal amount of system-induced autofluorescence, which is generated by the reflected laser light from internal connectors and crosstalk. A limitation of this approach is that this method does not work if sample fluorescence signals exist in all data acquired.

We consider the extracted autofluorescence spectrum, which may be caused by internal autofluorescence, to be the static background system autofluorescence. In the next step, we calculate the Pearson correlation coefficients (r value) between the background and all measured spectra to empirically estimate the sample fluorescence endmember. The spectrum with minimal r value is assumed to primarily consist of sample fluorescence. The precalculated background signal is subtracted from this spectrum to get the final sample fluorescence endmember for linear unmixing. An empirical threshold is chosen for r to ensure that sample fluorescence is sufficiently different to the system-induce autofluorescence.

After both $S_{em\_sys}(\lambda)$ and $S_{em}(\lambda)$ are calculated, we perform the least squares calculation based on boundary conditions: $\beta_{sys} \geq 1$ and $\beta_f \geq 0$. These constraints ensure that the sample fluorescence signal is non-negative and that the system autofluorescence components remain at or above the static background autofluorescence level. Upon finishing the calculation, we calculate the coefficient of determination ($R^2$) value between the observed value $P(\lambda)$ and reconstructed value $P_m(\lambda)$ to confirm the effectiveness of the algorithm. If the $R^2$ value is below an empirically selected threshold, the calculation will be considered as erroneous.

After processing with the algorithm, the final fluorescence signals will be reconstructed by applying the weighting factor $\beta_f$ to the emission spectrum $S_{em}(\lambda)$.

## 3. Imaging system and data processing

In current miniaturized fiber-optic fluorescence endoscopes, highly sensitive detectors such as photomultiplier tubes (PMTs) or avalanche photodiodes (APDs) are commonly used to detect fluorescence signals. However, for spectral unmixing, the emission fluorescence signals must be captured across different wavelengths to create a full spectrum. One method to achieve this is by sequentially using a set of emission filters, but this approach requires multiple images of the same location, which is time-consuming and impractical for many endoscopy applications. Alternatively, a spectrometer or a multi-anode PMT with a diffraction grating can be employed to simultaneously acquire the entire emission spectrum, enabling spectral unmixing without the need for repeated imaging. In our study, we opted for a high-sensitivity spectrometer to capture the full emission spectrum, facilitating efficient and accurate spectral unmixing of the fluorescence signals.

The system used for demonstrating our noise identification and removal methods is our previously published miniaturized single-fiber endoscope [26], where a DCF coupler (Castor Optics Inc., Canada) is used to integrate a 1300 nm optical coherence tomography (OCT) system (Telesto II, Thorlabs GmbH, Germany) and a spectrometer-based (QE Pro, Ocean Optics, USA) NIRF system. A custom DCF FORJ (Princetel Inc., USA) is used to couple both excitation and emission light between a stationary fiber and a rotating fiber probe. The excitation source is a fiber-coupled 785 nm laser (Integrated Optics, UAB), paired with a 785nm bandpass excitation filter (FBH785-3, Thorlabs Inc., USA). An 800nm long pass emission filter (FELH0800, Thorlabs Inc., USA) is used to block leaked excitation light. All the data were processed in MATLAB 2021b (The Mathworks, Inc., Natick, MA, USA).

The primary advantage of employing the OCT/NIRF multimodality system for algorithm validation lies in the ability of OCT to provide high-resolution structural information. This structural data allows for the visualization and identification of the hyperreflective artifacts in the OCT images. By correlating these artifacts with autofluorescence signals in the NIRF modality, we were able to verify the accuracy and stability of our algorithm, ensuring that autofluorescence artifact is effectively distinguished from true fluorescence signals.

**4. Phantom and in vivo validation**

We validated our algorithm first using a phantom setup with a mirror to generate system autofluorescence through reflection of the excitation light and a container filled with ICG solution (25g/ml in distilled water) to generate sample fluorescence signals.

Rodent *in vivo* studies [27] were further conducted to verify the effectiveness of our method. We analyzed 14 sets of mouse thoracic aortas *in vivo* from 5 mice. The NIRF agent ICG was injected into mice via the tail vein (130 μg/kg) one hour prior to the imaging. All experimental procedures were conducted with approval from the South Australian Health and Medical Research Institute Animal Ethics Committee (SAM 425.19) and conformed to the Guide for the Care and Use of Laboratory Animals (United States National Institute of Health).

**Results**

1. **Background noise reduction**

We measured background noise in our system under four configurations: 2 m long fiber patch cable without the bandpass filter, 2 m long fiber patch cable with the bandpass filter, 1 m long fiber patch cable without the bandpass filter, and 1 m long fiber patch cable with the bandpass filter. In Fig. 3a, results reveal that the addition of a bandpass filter decreased background noise intensity by about 40%, indicating that a significant portion of noise originates from leaked excitation light. Although a monochromatic laser typically would not require a filter in theory, laser wavelength broadening within the system introduces sideband noise to the measured fluorescence background [13]. After mitigating the excitation leakage with the excitation filter, the DCF remains the primary source of background autofluorescence, underscoring the impact of fiber length on autofluorescence levels. Reducing the connection fiber length from 2 m to 1 m led to a further 15% reduction in background noise.

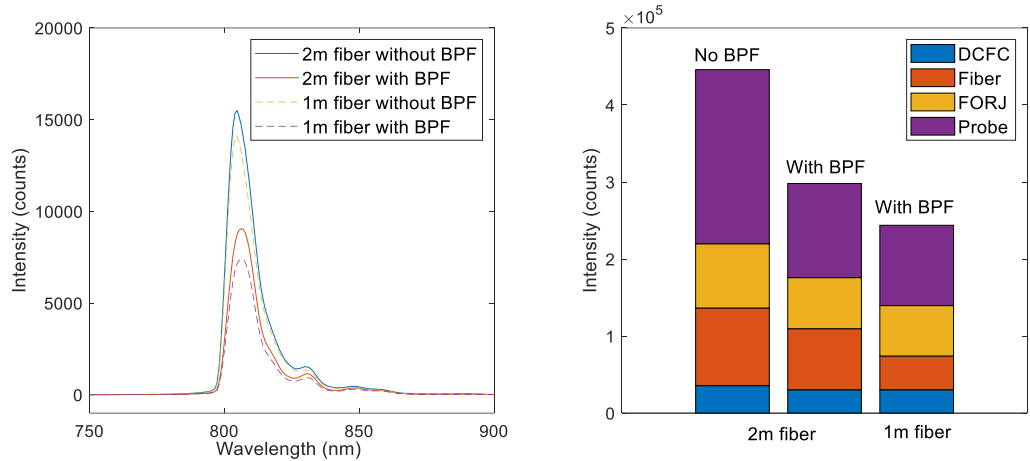

Fig. 3. Background noise reduction results. (a) Overall background noise measured under four different configurations: 2 m fiber patch cable without BPF, 2 m fiber patch cable with BPF, 1 m fiber patch cable without BPF, 1 m fiber patch cable with BPF. (b) Background intensities of four main components: DCF coupler, connection fiber, FORJ, and DCF probe. BPF, 785 nm bandpass filter; DCFC, double-clad fiber coupler; FORJ, fiber optic rotary joint.

To pinpoint noise sources, we assessed the background intensity in the four main system components: the DCF coupler, FORJ, connection fiber, and fiber probe. By removing each component sequentially, we isolated the background intensity of each component. As illustrated in Fig. 3b, without the bandpass filter, the fiber probe accounted for around 50% of the total background noise. Applying the excitation bandpass filter reduced noise across all components, with the most significant reduction at the fiber probe, likely due to reflection at the fiber/air interface. Reduced signal intensities in other parts also indicate reflection or crosstalk throughout the system. Moreover, while the fiber probe length must be sufficient to reach the coronary artery from a peripheral access point in our system, shortening the fiber patch cable between the DCF coupler and FORJ from 2 m to 1 m resulted in an approximately 45% reduction in autofluorescence in the connection fiber, as shown in Fig. 3b.

Collectively, these methods reduced the background noise level by half, enhancing the system dynamic range and sensitivities, resulting in improved detection of weak signals and imaging performance.

2. **Phantom test results**

Fig. 4 presents the results of a phantom test conducted to verify the performance of our method. Fig. 4a illustrates the phantom test setup, including a mirror to the left, and an ICG container to the right, while the DCF fiber probe moving horizontally from left to right. Fig. 4b presents the originally measured fluorescence signals, where both the mirror and the ICG appear to generate fluorescence signals in addition to the background. Fig. 4c demonstrates the fluorescence signals after static background subtraction, showing the system-induced autofluorescence generated by the reflected laser from the mirror still exists. Fig. 4d shows the effectiveness of our unmixing algorithm. The background and the artifacts from mirror reflection have been successfully removed, isolating the true fluorescence signal originating from the ICG. This indicates that our algorithm can accurately distinguish between true fluorescence and the autofluorescence artifacts caused by both internal and incidental external reflection.

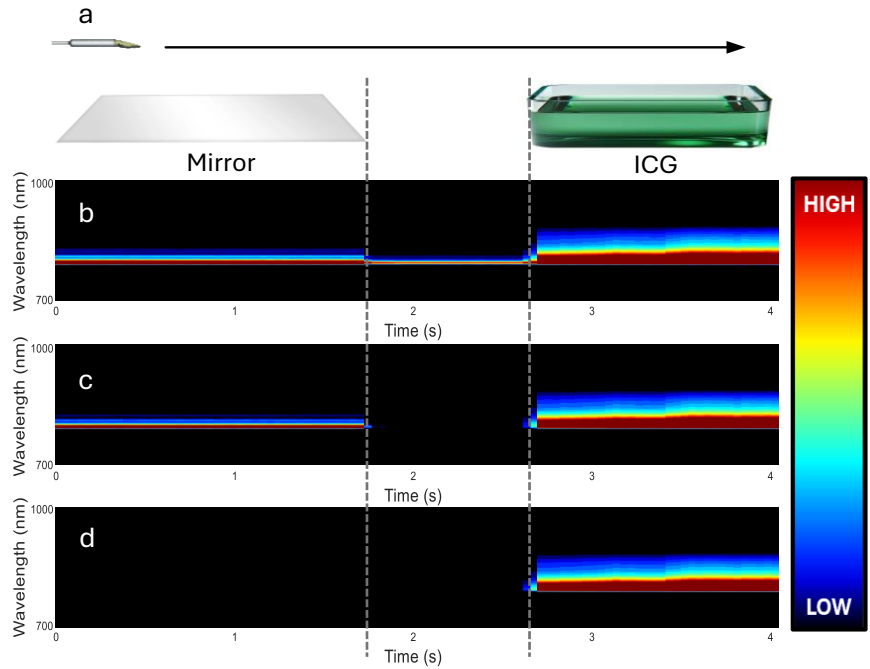

Fig. 4. Phantom test result. (a) Phantom and probe setup. (b) Original measured spectra. (c) Spectra after background subtraction. (d) Spectra processed with our unmixing method. ICG, indocyanine green solution.

### 3. Rodent *in vivo* study results

### 2.1. Removal of autofluorescence artifacts caused by external hyperreflective objects

Earlier studies on intravascular OCT/NIRF imaging have shown that enhanced ICG fluorescence signals effectively co-localize with lipid-rich, macrophage-abundant atheroma [28]. In this study, we validated the performance of our unmixing method by comparing the original and unmixed ICG fluorescence images with corresponding OCT images and histological analysis, ensuring accurate identification and removal of system-induced autofluorescence artifacts.

Fig. 5a shows a representative OCT/NIRF combined image highlighting macrophages (red arrows) and cholesterol crystals (CCs, green arrows). The OCT signal is displayed in gray in the center of the image, while the fluorescence signal is visualized as a thin, colored circle around the outside of the OCT signal. The original NIRF image reveals widespread fluorescence signals, with particularly strong signals at both sites and other locations. In Fig. 5b, constant background subtraction is applied to the NIRF signals, yet fluorescence remains visible at the same sites and additional areas. Fig. 5c presents the unmixed NIRF image and shows that fluorescence is confined exclusively to the macrophage regions. This is confirmed in Fig. 5d, which shows CD68$^+$ stained histology identifying macrophages (red arrows) at corresponding locations in the OCT/NIRF images. Fig. 5e represents hematoxylin and eosin (H&E)-stained histological cross-sections, indicating the presence of CCs (green arrows) at the same locations. Together, these images demonstrate that our unmixing method successfully removes system-induced autofluorescence artifacts, particularly at sites with CCs and other non-specific areas.

Similarly, Fig. 5f displays an OCT/NIRF combined image with prominent reflections from the catheter sheath (orange arrows). The original NIRF image shows widespread fluorescence and strong signals at the sheath reflection sites. After background subtraction (Fig. 5g), fluorescence persists at the reflection site. However, Fig. 5h shows the unmixed NIRF image, where no fluorescence is detected, indicating successful removal of the artifacts. This is further corroborated by Fig. 5i, which shows no CD68+ macrophage staining within plaques, and Fig. 5j, which provides H&E-stained cross-sections confirming the absence of specific targets such as CCs. Collectively, these results confirm that our unmixing approach effectively eliminates system-induced autofluorescence artifacts, including those caused by catheter sheath reflections.

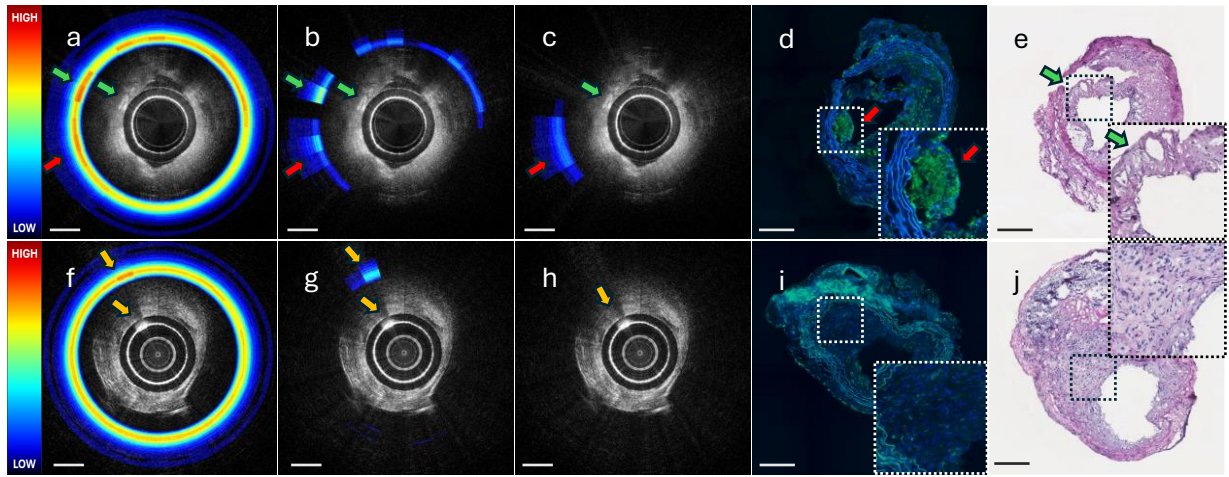

Fig. 5. Mouse *in vivo* unmixing result. Gray image in center of each figure shows the OCT image. The colored circle outside of the OCT image indicates the strength of the fluorescence signal. (a, f) OCT images combined with original measured NIRF images; (b, g) OCT images combined with background subtracted NIRF images; (c, h) OCT images combined with unmixed NIRF images; (d, i) Corresponding immunofluorescence images of cross sections showing CD68$^+$ macrophages (green staining, red arrow) combined with DAPI (4′,6-diamidino-2-phenylindole) staining; (e, j) Corresponding histological cross sections stained with H&E (hematoxylin and eosin); red arrows: macrophages; green arrows: cholesterol crystals; orange arrows: catheter sheath reflection. Scale bars: 250μm.

## 2.2. Correlation between fluorescence intensity and OCT background signal

Fig. 5b and 5g suggest that system-induced autofluorescence artifacts originate from external specular reflections, as identified as very high signal (white) in the OCT images (labelled with an arrow). To further substantiate this observation, we employed the $R^2$ value to examine the relationship between fluorescence intensity and OCT signal strength. Previous studies have shown that linear and planar reflection artifacts in OCT images, typically caused by hyperreflective objects, extend throughout the entire scan depth [29]. Therefore, we used OCT intensities from the image edges as background signals and compared these values with both the original fluorescence intensities and the target fluorescence signals extracted using our algorithm.

We analyzed the complete dataset from Fig. 5a, comprising 131,200 OCT A-scans and 2,624 fluorescence A-scans. Note that because the OCT was a higher resolution imaging modality,

50 OCT A-scans were acquired for each fluorescence measurement then averaged. The scatter plot Fig. 6 shows the correlation between OCT background signal levels and fluorescence intensity values. The original measured fluorescence values (blue circles) are highly correlated with the OCT background signal levels, with an $R^2$ of 0.5316 and a P-value of less than 0.0001. This strong positive correlation indicates that the initial fluorescence measurements are significantly influenced by the background signal in the OCT images, indicative of elastically backscattered signal. After processing with the algorithm, the extracted actual fluorescence values (red circles) show no obvious correlation with the OCT background signal levels, as evidenced by an $R^2$ of 0.0003 and a P-value of 0.4065. The lack of correlation indicates that the algorithm effectively eliminates the association with OCT background signals, which are typically indicative of external reflection strength. It is likely to result in more accurate and reliable fluorescence measurements that are not mixed with system-induced autofluorescence.

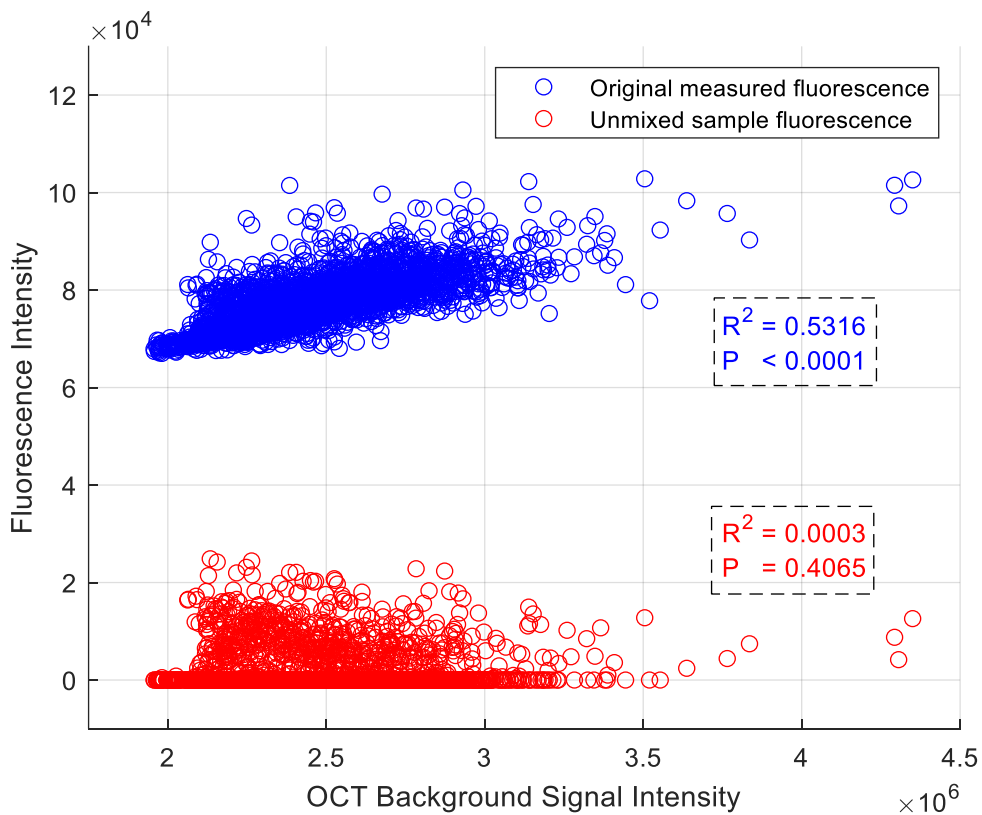

Fig. 6. Scatter plot of fluorescence intensity to OCT background signal intensity. Blue circles, original measured fluorescence value; Red circles, unmixed sample fluorescence. $R^2$, coefficient of determination; P, probability value.

## 2.3. Statistical performance of unmixing method

In addition to demonstrating the model's ability to separate autofluorescence artifacts from sample fluorescence, another useful validation lies in confirming the accuracy of the endmember spectra that represent the underlying components. To achieve this, we utilize the $R^2$ as a metric to quantify how well the variance in the observed spectrum $P(\lambda)$ is explained by the linear unmixing model. Specifically, in our case, the $R^2$ value measures how effectively the autofluorescence spectrum $P_a(\lambda)$ and the sample fluorescence spectrum $P_f(\lambda)$ account for the variability in the observed spectra. A high $R^2$ indicates that the separated endmember spectra closely match the original observed fluorescence, confirming that the separation process reliably preserves the true fluorescence signal and accurately isolates the autofluorescence component.

We analyzed all 14 *in vivo* mouse datasets and plotted the $R^2$ between the observed spectra $P(\lambda)$ and the reconstructed spectra $P_m(\lambda)$, with error bars representing the 99% confidence interval in Fig. 7. The average $R^2$ values across all datasets are consistently high, nearing one. This indicates an excellent agreement between the original and reconstructed spectra, demonstrating the robustness and accuracy of the unmixing method. The narrow error bars across the datasets further reinforce the reliability of the spectral reconstruction, as the 99% confidence intervals show minimal variability, underscoring the precision of the algorithm in all scans.

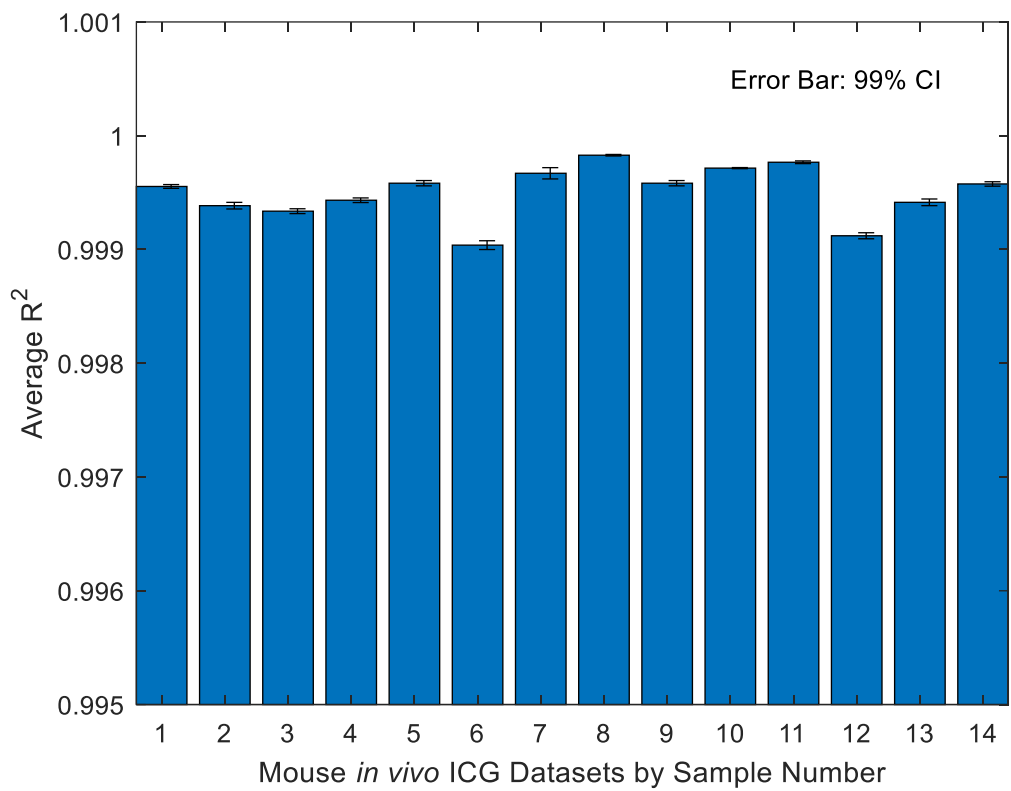

Fig. 7. Average coefficient of determination ($R^2$) between original spectra and reconstructed spectra from unmixed components with error bar showing 99% confidence interval.

**Discussion and conclusion**

In this study, we discussed the sources of the system-induced autofluorescence, provided the optimization methods to reduce the background noise floor, and developed a linear unmixing method to effectively unmix system-induced autofluorescence from sample fluorescence signals acquired with a miniaturized fiber-optic fluorescence endoscope. The method was validated through a phantom test and *in vivo* mouse experiments, demonstrating its accuracy and reliability in distinguishing autofluorescence artifacts from true fluorescence signals. The results confirm the potential of this approach to improve fluorescence imaging, particularly in contexts where system-induced autofluorescence presents a significant noise to the measurement.

In addition to successful unmixing, several other outcomes are possible after processing with our algorithm. First, if no sample fluorescence is present, the correlation coefficient r between the extracted autofluorescence and the supposed sample fluorescence signal will typically be large and can be detected by setting an empirical threshold. Such instances indicate that no sample fluorescence signal exists. Second, in the absence of external artifacts and changing background noise floor, we found that the algorithm's result will closely resemble that of simple background subtraction. Note that if the $R^2$ value between the original measured spectra and the reconstructed spectra is too low, the algorithm will not succeed. This failure could be due to the presence of additional sources of fluorescence, such as significant tissue autofluorescence, or external noise within the emission spectrum range. In our studies, tissue autofluorescence at NIR range is negligible compared to the ICG fluorescence, and as a result, we did not experience any algorithm failure caused by the presences of additional source of fluorescence.

This proposed method has the potential to offer a robust and adaptable tool for enhancing signal separation across various biomedical imaging applications. In our experiments, we used a single-fiber OCT/NIRF multimodality endoscope to measure the ICG induced fluorescence signals. Although single-fiber system is more vulnerable to system-induced autofluorescence due to the common path of the emitted excitation light and detected fluorescence emission, autofluorescence artifacts caused by external specular reflections can affect both single-fiber and multi-fiber endoscopes. This broadens the applicability of our method, making it potentially useful for a wider range of fiber-based endoscopy setups. Similarly, the algorithm we developed is adaptable to other fluorescent agents (not limited to ICG), provided that the system's autofluorescence spectrum is distinct from that of the fluorescent agent. Moreover, the algorithm is also effective in near infrared autofluorescence imaging (NIRAF) [7, 30-33], if the tissue autofluorescence spectrum is independent of the system autofluorescence spectrum. This versatility underscores the broader applicability of our proposed method.

The background autofluorescence of an imaging system (not induced by the sample) is often referred to as static noise, but it can fluctuate under certain conditions. Fluctuations in the laser source stability may alter the intensity of background autofluorescence. In most miniaturized fluorescence endoscopes, a FORJ is used to couple light between a stationary fiber and a rotating fiber. Variations in coupling efficiency during rotation, influenced by alignment accuracy and motor stability, can lead to changes in background noise levels. These factors make simple background subtraction insufficient. However, our method effectively addresses these challenges by not assuming a constant value of system background autofluorescence.

One limitation of our implementation is that it required a spectrometer to capture the full emission spectra, which allowed for precise spectral unmixing. However, for high-speed clinical applications, a multi-anode PMT combined with a diffraction grating [34, 35] may be a more cost-effective solution. The choice between these configurations must be carefully considered, balancing clinical application requirements, cost constraints, and performance needs.

**Funding**

This work is supported by National Health and Medical Research Council (NHMRC) Ideas Grant 2001646 (JL, RAM, CAB), NHMRC Development grant 2022337 (JL, RAM), NHMRC Investigator Grant 2008462 (JL), Heart Foundation Future Leader Fellowship 105608 (JL), Hospital Research Foundation project grant 2022-CP-IDMH-014-83100 (JL, RAM), NHMRC Ideas grant 1184571 (CAB), Lin Huddleston Heart Foundation Fellowship (CAB).

**Data Availability**

The data supporting the finding of this study and the codes to reproduce the results of this article are available from the corresponding author upon reasonable request.